# Tunnel Junction-Enabled Monolithically Integrated GaN Micro-Light Emitting Transistor

Sheikh Ifatur Rahman[1], Mohammad Awwad[1], Chandan Joishi[1], Zane Jamal-Eddine[1], Brendan Gunning[2], Andrew Armstrong[2], and Siddharth Rajan[1,3]
[1]Department of Electrical and Computer Engineering, The Ohio State University, Columbus, Ohio, USA.
[2]Sandia National Laboratory, Albuquerque, New Mexico, USA.
[3]Department of Material Science and Engineering, The Ohio State University, Columbus, Ohio, USA.
Email: rahman.227@osu.edu

**Abstract**: GaN/InGaN microLEDs are a very promising technology for next generation displays. Switching control transistors and their integration are key components in achieving high-performance, efficient displays. Monolithic integration of microLEDs with GaN switching devices provides an opportunity to control microLED output power with capacitive (voltage) control rather than current controlled schemes. This approach can greatly reduce system complexity for the driver circuit arrays while maintaining device opto-electronic performance. In this work, we demonstrate a 3-terminal GaN micro-light emitting transistor that combines a GaN/InGaN blue tunneling-based microLED with a GaN n-channel FET. The integrated device exhibits excellent gate control, drain current control and optical emission control. This work provides a promising pathway for future monolithic integration of GaN FETs with microLED to enable fast switching high efficiency microLED display and communication systems.
**Key Terms**— microLED, tunnel junction, transistor, integration, display, monolithic

Gallium Nitride-based micro-light-emitting diodes (microLEDs) are the key technology enabling emerging next-generation microLED displays for wearable devices, phones, and AR/VR applications and have potential applications in Li-Fi and biomedical sensing. In such portable display applications, capacitive or voltage-controlled methods to regulate LED arrays can significantly reduce complexity when compared with current-controlled methods. In recent years, microLEDs have been integrated with thin film transistors (TFTs) heterogeneously[1-7] or monolithically[8-17] to create a voltage-controlled scheme. However, LEDs with integrated switching transistors (i.e., III-Nitride based LEDs and transistors) can take advantage of high-performance GaN field effect transistors. Such integrated 3-terminal LED-transistors have the potential to reduce switching charge, increasing switching speed, and reduce system complexity when compared with microLED arrays that are directly driven by TFT/CMOS-circuits. In this work, we demonstrate a micro light-emitting transistor (**LET**) with an n-channel GaN field effect transistor in series with a GaN LED.

Figure 1(a) shows the individual components of the light-emitting transistor, and the corresponding equivalent circuit . The 3-terminal micro light-emitting transistor (**LET**) consists of three component devices that are epitaxially integrated, from bottom to top: a conventional III-Nitride LED, an interband tunnel junction, and an n-channel field effect transistor (control transistor). The top n-channel control transistor provides gate control of the current into the tunnel junction-LED. A buried insulated region is created to isolate the field effect transistor region from the LED.

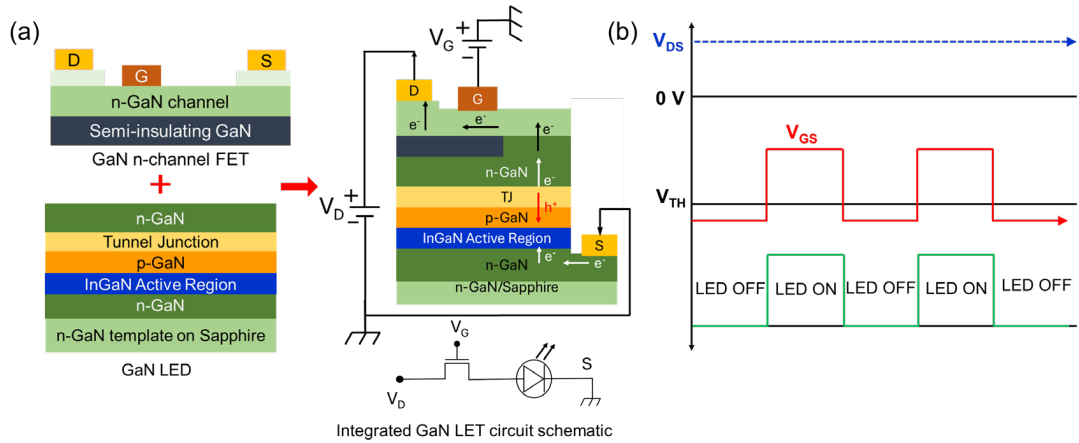

**Figure 1.** (a) Integration of GaN n-channel FET and GaN LED to develop three terminal GaN micro-Light emitting transistor (b) Timing diagram of the operation of GaN LET

The operating principle of the device is similar to that of a TJ-based LED[18-24] except with a gated structure at the top to modulate the charge. Under operation, the top (drain) contact enables tunneling-based hole injection to the active region through the p-GaN while electrons are injected through the bottom contact (in this case, source contact). The gate controls the charge modulation in n-channel FET region, which enables modulation of the current into the top tunnel junction. The timing diagram in figure 1(b) shows the operating state of this light emitting transistor. The key advantage of a three-terminal gated LED is that it can greatly reduce the requirements on the switching charge for driving LEDs. In case of the conventional two-terminal PN junction LEDs, the charge accumulated on the diode capacitance is relatively large since it includes the depletion and minority carrier stored charge, and a high driving current is needed to modulate this charge under switching conditions[25]. However, in the gated LED structure, the device is switched by only modulating the gate charge ($Q_G$), which can be designed to be 20× lower than the PN junction switching charge. Previous TJ-based approaches used vertical side-wall control of the devices using FinFET[26] or a vertical trench MOSFET[27] structure. Both these approaches are somewhat challenging in terms of manufacturability. The FinFET approach requires high aspect ratio sub-micron structures that are difficult in conventional LED manufacturing, while the trench MOSFET has a dielectric-semiconductor interface that poses challenges related to carrier transport (low inversion layer mobility) and dielectric-semiconductor interface traps. The approach proposed here is compatible with conventional LED processing and gives higher transistor mobility and greater control of the gate charge by designing the channel doping and thickness.

The device was fabricated using metal-organic chemical vapor deposition (MOCVD) to grow the LED and the continuously grown TJ at the top. The GaN-based blue LED (including TJ) was grown on a n-GaN on sapphire template. The LED active region consisted of 3 pairs of GaN/InGaN barrier/quantum well pairs. The homojunction TJ consisted of 12 nm heavily doped $p^{++}$ GaN and 10 nm heavily doped $n^{++}$ GaN, followed by thick n-type GaN current spreading layer. To integrate the gated structure at the top, a semi-insulating GaN layer is required to block the vertical current underneath the drain. The highly resistive semi-insulating layer was created using nitrogen ion (N) implantation technique with the dose conditions of 25 keV, $2 \times 10^{14}$ ions/cm² and 120 keV, $6.50 \times 10^{14}$ ions/cm² with a depth target of around 200 nm. A similar implant condition was previously shown to provide good isolation for GaN/AlGaN HEMT on Si[28]. The implant creates a current aperture to establish a series electrical connection between the n-channel FET gated channel and the LED. The dimension of the implanted area can be optimized based on the need of the gated structure and microLED dimensions. The p-GaN layer of the sample was thermally activated using rapid thermal anneal in $N_2$ ambient at 900 °C for 15 minutes[21, 29]. The sample was cleaned using HCl (37%) solution and organic solvents (Acetone/Methanol/IPA for 5 mins with ultrasonication). The sample was then loaded into the molecular beam epitaxy (MBE) chamber for the channel layer regrowth. The channel layer was designed to have ~ 80 nm of lightly doped GaN with Si concentration of $1 \times 10^{18} cm^{-3}$. The regrowth was done at substrate temperature ($T_{SUB}$) of 750 °C under Ga-rich growth conditions to ensure high quality GaN growth with a nominal growth rate of 5 nm/min. The top of the channel was capped with 25 nm $n^{++}$-GaN for ohmic contact.

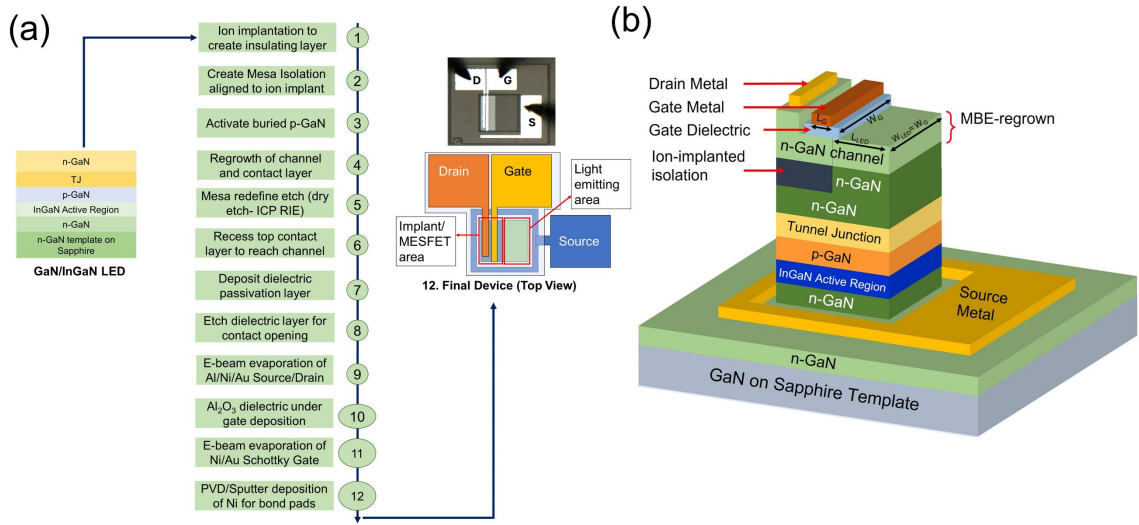

**Figure 2.** (a) Detailed process/fabrication steps for GaN light emitting transistor fabrication (b) 3-D schematic of the final device.

The detailed steps of the fabrication process and final device schematic are shown in figure 2(a). All the lithography steps were done using direct write optical lithography (Heidelberg MLA150). The device mesa isolation was done using inductively coupled plasma-reactive ion etching (ICP-RIE) using $BCl_3/Cl_2/Ar$ chemistry. On the gate area, the top 25 nm $n^{++}$-GaN was recessed to reach the channel layer using similar etch chemistry. After the mesa isolation and gate recess process, the sample was cleaned in 80°C tetramethylammonium hydroxide (TMAH), buffered oxide etchant (BOE) and hydrochloric acid (HCl) to remove any etch residues from the surface/sidewalls. $SiO_2$ was then deposited using plasma enhanced chemical vapor deposition (PECVD) technique for passivation and to serve as dielectric separation for bond pads. $SiO_2$ was etched from the drain, gate and source contact area using a combination of dry and wet etch conditions. An Al (30 nm)/Ni (30 nm)/Au (100 nm) unannealed ohmic contact was co-deposited using electron-beam evaporation. To ensure low gate leakage and higher forward blocking, $Al_2O_3$ was deposited using thermal atomic layer deposition (ALD) technique. The thickness of the deposited $Al_2O_3$ was measured as 11 nm using ellipsometer on a Si piece coloaded with the sample. Ni (20 nm)/Au (80 nm) Schottky gate contact metal was deposited using electron-beam evaporation technique.

We first characterized the regrown channel layer charge profile using capacitance-voltage (C-V) measurement technique. The structure used for C-V measurement to probe channel charge profile is shown in figure 3(a) along with different possible capacitances in the structure. The equilibrium band diagram of the layers underneath the Schottky Ni/Au gate (11 nm $Al_2O_3$/ 80 nm n-GaN ($1x10^{18}$ $cm^{-3}$, Si)/semi-insulating implanted GaN/doped GaN) is shown in figure 3(b). Figure 3 (c) shows the measured capacitance underneath the gate structure and the simulated capacitance profile. The two terminal capacitance profile extracted from Silvaco TCAD matches with the measured capacitance suggesting that the experimental structure matches the design shown in figure 3(a). The sheet charge density in the simulated structure is ~ $4.5 \times 10^{12}$ cm$^{-2}$. Figure 3(d) shows low two-terminal gate to source/drain current.

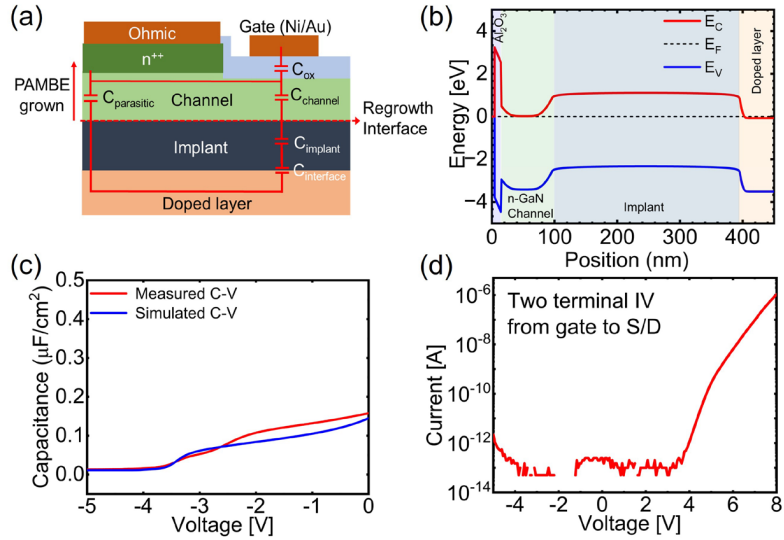

**Figure 3.** (a) Structure used for capacitance-voltage profiling and all the capacitances in the measured structure, (b) simulated equilibrium energy band diagram (SILVACO) (c) capacitance-voltage profile (measured and simulated), and (d) two terminal I-V characteristics from gate to source/drain.

We also investigated the performance of the standalone n-channel FET structure on the implanted semi-insulating surface which is normally on device. The device structure of the measured n-channel FET is shown in figure 4(a). Figure 4(b) shows the extrinsic transconductance measurement (measured for drain bias, $V_{DS}$ at 10 V) of the standalone n-channel FET with gate length, $L_G = 2$ μm and gate width, $W_G = 50$ μm on implanted layer. The device shows peak transconductance, $g_m$ of 50 mS/mm with peak source-

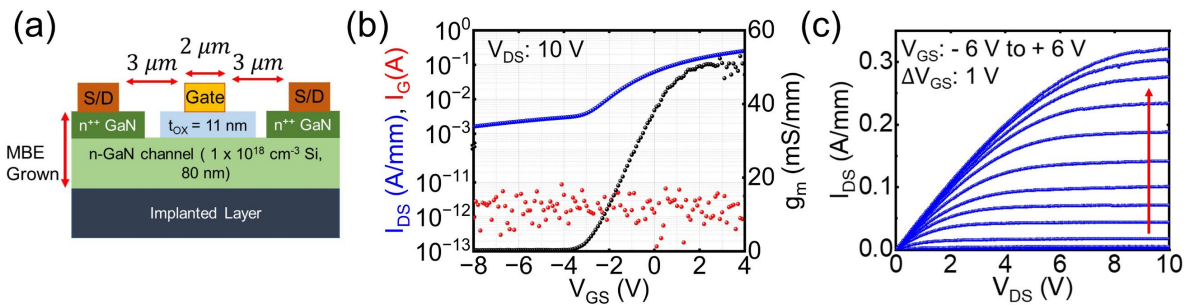

**Figure 4.** (a) Schematic of a stand-alone n-channel FET, (b) transfer characteristics of the stand-alone n-channel FET, (c) output characteristics of the n-channel FET.

drain current, $I_{DS}$ of 250 mA/mm at gate bias $V_G = +4$ V. Figure 4 (c) shows the $I_{DS} - V_{DS}$ characteristics of the device at room temperature for gate bias from $-6$ V to $+6$ V with $\Delta V_{GS} = 1$ V. An $I_{DS}$ of 321 mA/mm was measured at $V_{DS} = 10$ V for a gate bias of $+6$ V. The device $R_{ON}$ is estimated to be

18.5 Ω-mm at $V_{GS} = +6$ V. The threshold voltage was $-4$ V, which matches the estimate from C-V measurements.

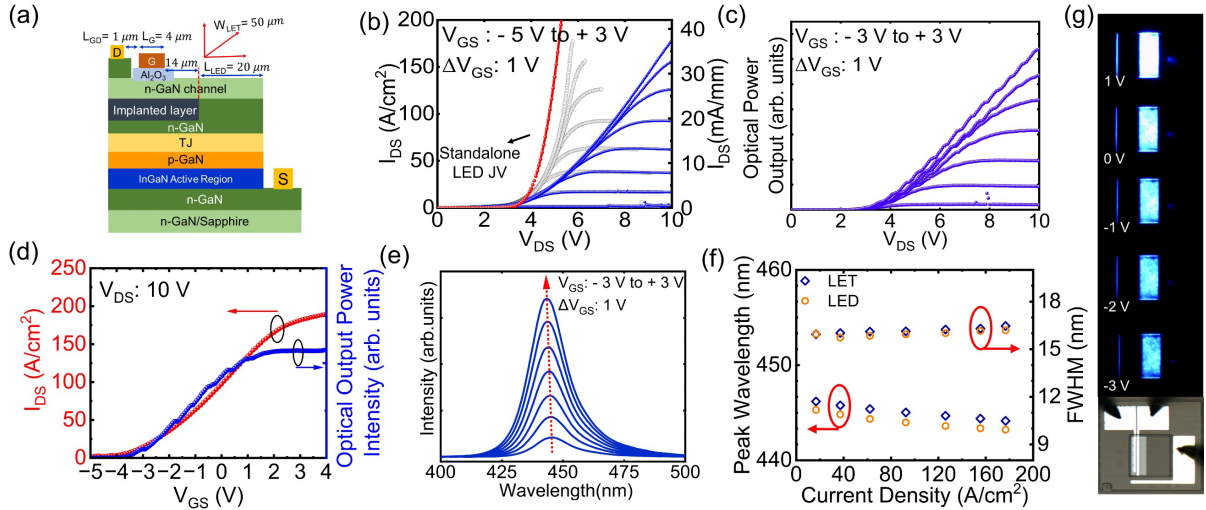

**Figure 5.** (a) Schematic of the measured device (b) Measured output current ($I_{DS}$-$V_{DS}$-$V_{DS}$) characteristics, (c) measured output light (L-$V_{DS}$-$V_{GS}$) characteristics, (d) transfer characteristics, (e) gate bias-dependent electroluminescence spectrum, (f) EL peak shift and full-width half maximum for LET and standalone microLED with TJ, and (g) optical micrograph of the light-emitting transistor shown in (a).

We report here on micro-light emitting transistors with emission area of 20 μm × 50 μm (defined by ion implantation). The gated structure on the top with gate length, $L_G$ of 4 μm and gate width, $W_G$ of 50 μm. The drain to gate spacing was 1 μm. Based on optical micrograph images, we find that the region under the implant did not have any emission, most likely due to a resistive n-region below that prevented current spreading. Therefore, the output characteristics of the n-channel FET /microLED were normalized to the dimension of the microLED injection/emission area.

Figure 5(a) shows the schematic of the measured/reported device. The $I_{DS} - V_{DS}$ measurements of the light emitting transistor measured at room temperature is shown in figure 5 (b) . The device turns on for a gate bias above $-4$ V and the turn on voltage of the device corresponds to the total turn on voltage drop of the microLED and the TJ. The total forward voltage drop of the device includes the voltage drop from LED (PN junction), TJ, and channel of the gated structure. The spacing between the implanted region edge and source end of the gate adds resistance that degrades transconductance and on-resistance of the control transistor. For example, the operating voltage for the LET (at 100 A/cm$^2$, $V_{GS} = +3$ V) is 2.95 V higher than that of the standalone LED. The sheet resistance of the regrown channel layer is estimated to be 5.65 kΩ using four terminal transfer length measurements. For the device with dimensions shown in Figure 5(a), at 100 A/cm$^2$, the voltage drop between the drain metal edge and the implant edge (~ 19 μm) can therefore be estimated to be ~ 2.5 V, indicating that the majority of the extra voltage drop for LET operation is due to this region. The de-embedded IV curves after removing this resistance are shown in Figure 5(b) (in black open circles). Further scaling and optimization of the channel charge profile can help to reduce this voltage.

The optical power response characteristics, as expected, are a replica of the output drain current characteristics. Figure 5(c) shows the optical output power (in arbitrary units) for the corresponding $I_D - V_D$ measurements shown in figure 5(b). The output power (from $-3$ V to $+3$ V with 1 V increment) shows good modulation of optical power with the current modulation through the device. Figure 5(d) shows the

$I_{DS} - V_{GS}$ characteristics with corresponding optical power response from $-5$ V to $+4$ V for gate bias measured at $V_{DS}$ of 10 V. Figure 5(e) shows the electroluminescence measurement of the device from gate bias of $-3$ V to $+3$ V with 1 V increment at $V_{DS} = 10$ V. The peak emission wavelength of the device at low current density ($V_{GS} = -3$ V) is 446 nm which shifts to 443 nm at high current density ($V_{GS} = +3$ V). The optical modulation enabled by the integration of the gated structure at the top shows that the microLED emission intensity can be controlled efficiently as a voltage-controlled device rather than a current controlled device. Figure 5(f) shows the full width half maximum and peak wavelength shift of the electroluminescence spectra from the LET and standalone LED operating under similar current injection. Both LET and standalone LED show nearly unchanged full width half maximum and similar peak wavelength shift with varying current density which indicates that there is no significant degradation or alteration of emission characteristics of the active region during the implantation and fabrication process. Figure 5(g) shows the optical micrograph of the device under different gate bias showing uniform emission from the device under all current injection (gate bias). The optical micrograph shows emission only in the non-implanted regions, confirming that the implanted region blocks vertical current, and also that there is no spreading of the current underneath this region. Above $+1$ V of $V_G$ the camera used in this setup was saturated by the emission. The device showed uniform emission indicating uniform current injection through the tunnel junction.

In conclusion, in this work we designed and demonstrated a monolithic integration process for GaN microLED and GaN FET capitalizing on the advantages of a TJ-based contact for the LED. The micro-light-emitting transistor developed here utilizes the highly conductive n-GaN at the top of the tunnel junction to create a gate structure where the charge is controlled. This device shows exhibit excellent gate control and charge modulation with optical output modulation similar to the microLED. Further research may lead to growing the entire structure using the MOCVD process. Future work could include studying alternate structures as well as methods to achieve the top channel implantation. This could include MOCVD-based regrowth processes, AlGaN/GaN-based HEMT channels for higher mobility, and patterned doping using ion-implantation. Excellent gate control and low charge in the channel demonstrated in this work could enable low power consumption and faster switching for the microLED, enhancing its performance in display and communication applications. Moreover, this monolithic integration process can be applied to TJ-based microLED for any wavelength, any number of active regions, or any orientation of the PN junction of the light emitting diode.

This material is based upon the work supported by the U.S. Department of Energy's Office of Energy Efficiency and Renewable Energy (EERE) under the Building Technologies Office award No. 31150, ONR Grant No. N00014-22-1-2260 and INTEL-CAFE Project. This article has been authored by an employee of National Technology & Engineering Solutions of Sandia, LLC under Contract No. DE-NA0003525 with the U.S. Department of Energy (DOE). The employee owns all right, title, and interest in and to the article and is solely responsible for its contents. The United States Government retains and the publisher, by accepting the article for publication, acknowledges that the United States Government retains a non-exclusive, paid-up, irrevocable, world-wide license to publish or reproduce the published form of this article or allow others to do so, for United States Government purposes. The DOE will provide public access to these results of federally sponsored research in accordance with the DOE Public Access Plan https://www.energy.gov/downloads/doe-public-access-plan. This paper describes objective technical results and analysis. Any subjective views or opinions that might be expressed in the paper do not necessarily represent the views of the U.S. Department of Energy or the United States Government.